\documentstyle[12pt]{article}
\def \ni {\noindent}

\def \o  {\overline}
\def \ra {\rightarrow}
\def \f  {\frac}
\def \non {\nonumber}

\def \lc {{\cal L}}
\def \ba {\begin{array}}
\def \ea {\end{array}}
\def \be {\begin{equation}}
\def \ee {\end{equation}}
\def \bea{\begin{eqnarray}}
\def \eea{\end{eqnarray}}

\def \idt {{\rm Im}\,d_t}
\def \rdt {{\rm Re}\,d_t}

\def \bt{\beta_t}
\def \bg{\beta_{\gamma}}

\def \om {\omega}
\def \sg  {\sigma}
\def \la{\lambda}

\def \lae {\lambda_e}

\def \lg {\lambda_{\g}}
\def \ll {\lambda_l}
\def \lt {\lambda_t}
\def \ltb {\lambda_{\o t}}
\def \tl{\theta_l}
\def \tht{\theta_t}
\def \th0{\theta_0}

\def \l  {Lagrangian}
\def \cp {$\boldmath CP$}

\def \g {\gamma}

\def \gg {\g \g}

\def \tbar {$\o t$}

\def \ttbar{$t \o t$}

\def \ep {$e^+ e^-$}
\def \eett {$e^+ e^- \ra t \o t$}
\def \ggtt {$\g \g \ra t \o t$}

\def \ach {$A_{ch}$}
\def \afb {$A_{fb}$}

\title{
Electric Dipole Moment of Top Quark and $CP$-Violating Asymmetries in 
$\gamma \gamma \rightarrow t \overline{t}$
}
\author{
P. Poulose and Saurabh D. Rindani\\[3mm]
{\it Theory Group, Physical Research Laboratory,}\\
{\it Ahmedabad - 380 009, India}
}
\date{}

\begin{document}
\maketitle
\begin{abstract}
\cp-violating asymmetries due to a possible electric dipole interaction of the
top quark in the production and subsequent decay of top
quark-top antiquark pair in photon-photon collisions are studied.
The asymmetries 
defined can be used to determine the imaginary part of the electric 
dipole form factors.  A $\gamma \gamma$
collider with photon beams generated from laser back-scattering off 
electron beams with an
integrated geometric luminosity of $20\,fb^{-1}$ can put
a limit of the order of $10^{-17}\;e\,cm$ on the imaginary part of the
electric dipole form factor of the top quark if 
the electron beams have longitudinal polarization and the laser beams have circular
polarization.
\end{abstract}
\vskip .2cm


\section{Introduction}
Studying \cp\ violation and looking for its signatures at colliders
(present and future) is important for  
various reasons.
Apart from the fact that any indication of \cp\ violation outside the 
K-meson and B-meson systems seen at 
colliders likely to be operational in the near future  will be a
clear indication of physics beyond the standard model (SM), the 
phenomenon of \cp\ violation requires to be understood in greater detail.
With the expectation and the eventual discovery of the heavy top quark,
\cite{expt} there have been many studies on the 
electric/weak dipole form factor 
of the top quark and the \cp\ violation it induces 
\cite{model}-\cite{hadron}.  
This includes work on signatures of \cp\ violation in
\ttbar\ production at \ep\ colliders 
\cite{gzad}-\cite{paul2},  
as well as at hadron colliders \cite{hadron}.

Since synchrotron radiation makes
increasing the energy of a circular collider far beyond that of LEP2
prohibitive, future colliders operating at higher energies will have to be 
linear colliders. In the
context of linear colliders, the possibility of photon linear colliders
has been discussed in the literature \cite{ginzburg,ginzburg2}.  
The hope is that such a
collider will be operational in the coming decade.  In such colliders an
intense low-energy laser beam would be backscattered by a high energy
$e^+/e^-$ beam to give a high-energy photon beam.  This photon beam could then
be made to collide with another photon beam or with
another lepton beam.  In this article we discuss signals of possible 
$CP$ violation in the production of 
a top quark - top antiquark pair in a photon-photon collider, by examining
asymmetries in angular distribution of leptons (anti-leptons) that arise in a
semi-leptonic decay of the $t\overline{t}$ pair.

The topic of \cp\ violation in \ggtt\ has also elicited interest recently.
Anlauf  {\it et al.} \cite{anlauf} and Bernreuther {\it et al.} \cite{berngg} 
have discussed \cp\ violation in 
a Higgs mediated \ggtt\ process where they study triple-product correlations
as well as asymmetries.  Choi and Hagiwara \cite{choigg}, and Baek {\it et al.}
\cite{Baek} have studied the
effect of the top quark electric dipole form factor (EDFF)
in asymmetries in the top distribution in \ggtt\ with linearly 
polarized photon beams.

Here we assume a $CP$-violating electric-dipole interaction of the top quark, 
but neglect $CP$ violation in top decay.
We consider two asymmetries: the charge asymmetry, which is the
asymmetry in the number of leptons and antileptons produced in the decay of
the top antiquark and the top quark with a \cp -conserving angular cut in the
forward and backward directions, and the charge asymmetry 
combined with the forward-backward asymmetry. In the absence of a cut-off,
charge asymmetry becomes the asymmetry in the production rates of 
$t$ and \tbar,
which is zero due to charge conservation in the absence of $CP$ violation in 
top decay. The effect of longitudinal electron beam polarization together
with circular laser beam polarization on the asymmetries
is also studied.
With an integrated geometric 
luminosity of 20 fb$^{-1}$ and an initial electron beam energy of a few 
hundreds of GeV, the limit that can be placed on the imaginary part of 
the top quark EDFF is found to be 
of the order of 
$10^{-17}\,e\,cm$. 

The paper is organized as follows. Main features of a photon linear collider 
will be discussed in the next 
section (Section 2). In Section 3 we derive expressions for the $CP$-violating 
asymmetries. The results and conclusion are contained in Section 4.

\section{Features of a $\gg$ Collider}

\begin{figure}
\vskip 6cm
\includegraphics{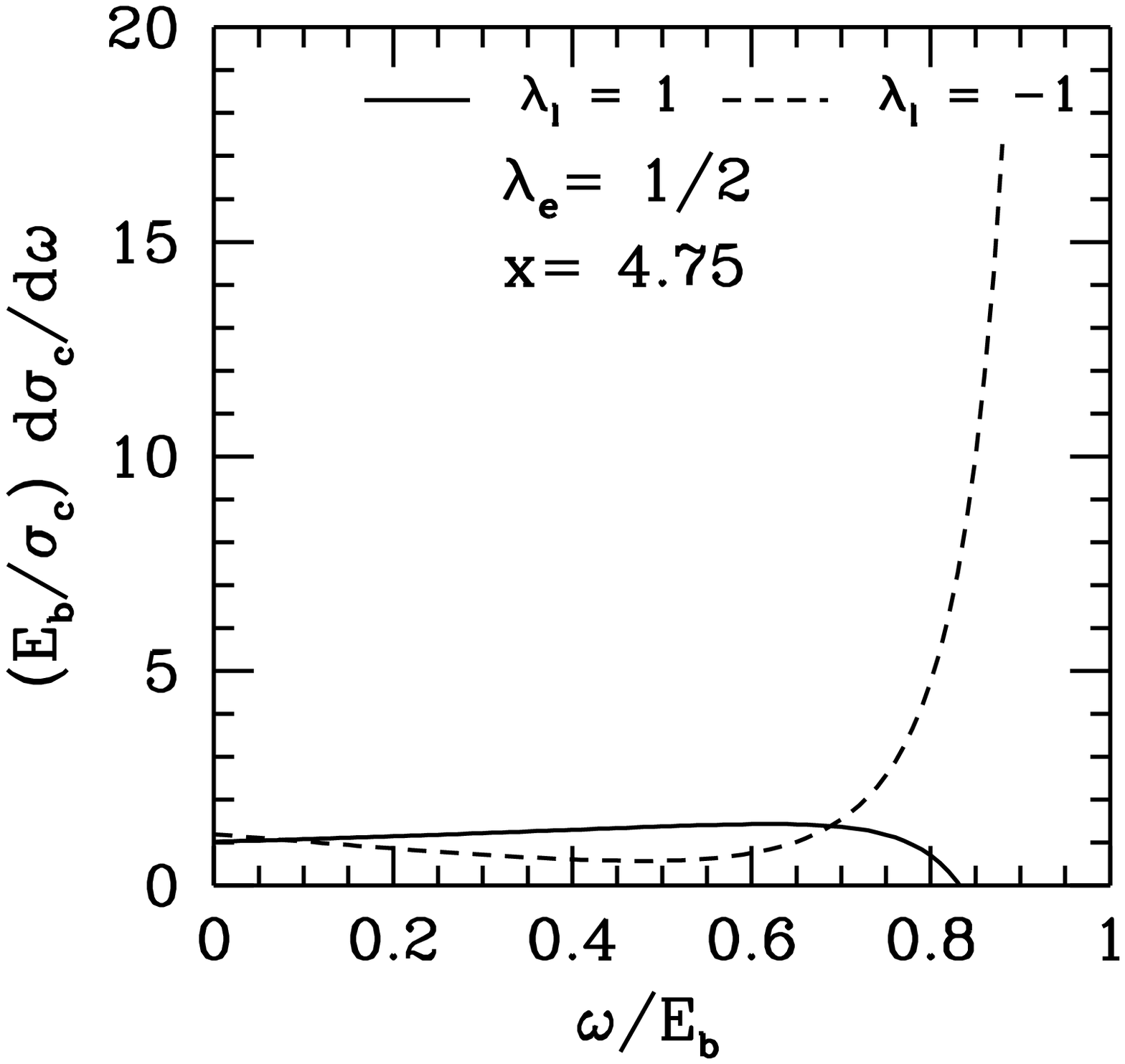}
\includegraphics{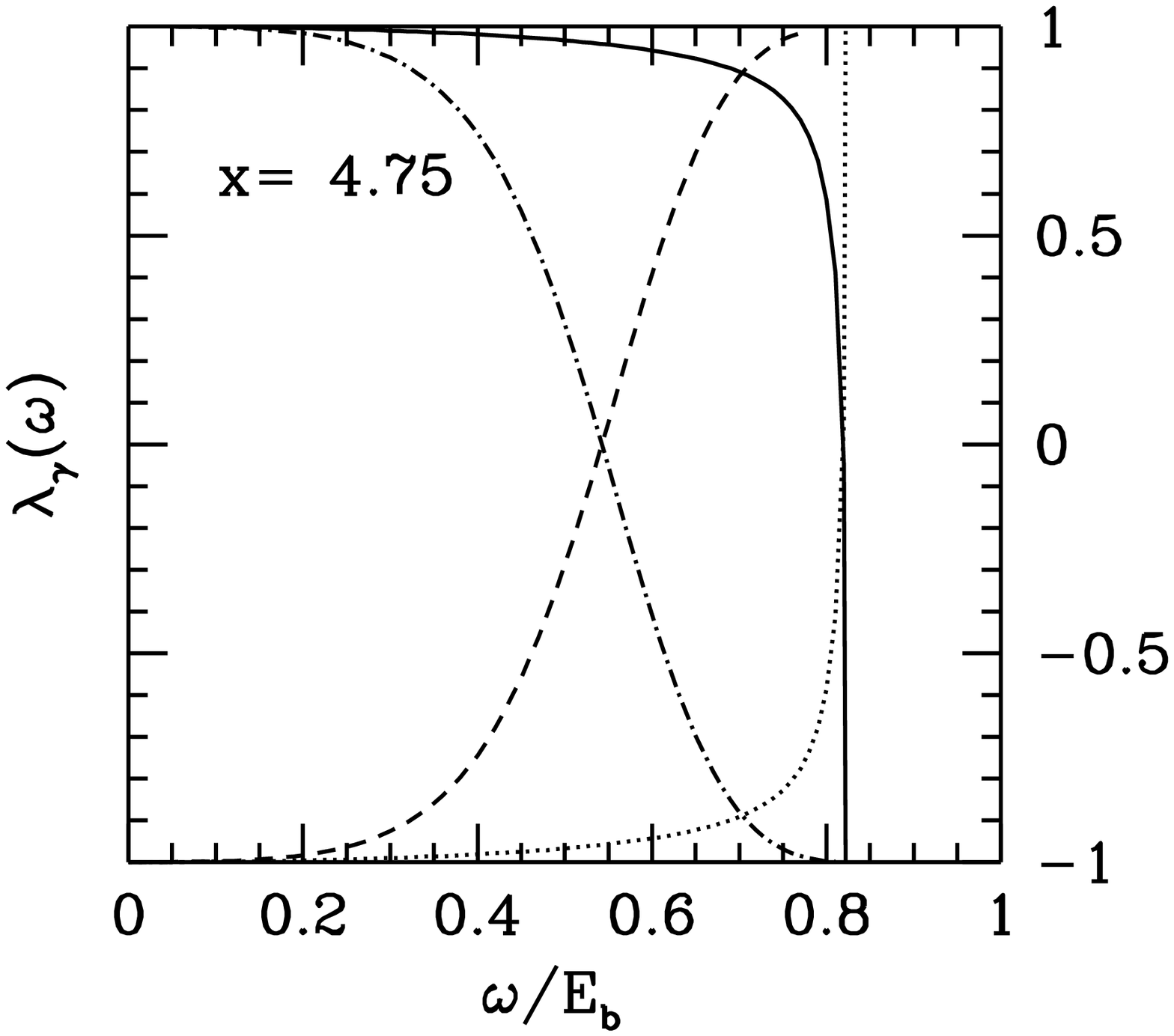}
\caption[dummy]
{\small The figure on the left shows the energy distribution of 
Compton-scattered photons
for
different helicity combinations of the initial electron beam and the laser
beam. In the figure on the right the scattered photon helicity is
plotted against the
energy of the scattered photon for different helicity combinations of 
the laser beam and the initial electron beam. Solid and dotted lines 
correspond to $2\,\lae \ll=1$ with $\lae=1/2$ and $\lae=-1/2$ respectively
while the dashed line and the dash-dotted line correspond to $2\,\lae 
\ll=-1$ with $\lae=1/2$ and $\lae=-1/2$, respectively.}
\label{fig:dist_pol}
\end{figure}

	In a $\gg$ collider, high-energy photons would be produced 
by Compton backscattering of
intense low-energy laser beams off high energy electrons \cite{ginzburg}.
The energy spectrum of a Compton-scattered photon is given by
\bea
\f{1}{\sigma_c}\f{d\sigma_c}{dy} &=& f(x,y) \non\\
&=&\f{2 \pi \alpha^2}{\sigma_c
x m_e^2}\left[\f{1}{1-y}+1-y-4r(1-x)-2\lambda_e
\lambda_{l}rx(2r-1)(2-y)\right]. 
\label{eq:endist}
\eea
Here
\bea
 x=\f{4E_b\omega_0}{m_e^2}=15.3\,\left(\f{E_b}{\rm TeV}\right)
\,\left(\f{\om_0}{\rm eV}\right),
\eea
where $E_b$ is the electron beam energy, $\om_0$ is the
energy of the laser beam and $m_e$ is the electron mass. 
$y$ is given in terms of the energy of the scattered photon, \(\om\;
(\leq E_b\f{x}{1+x}) \) as
\[y=\f{\om}{E_b}\] and \[r=\f{y}{x (1-y)}\leq 1.\] 
$\lae$ and $\ll$ are the initial electron and laser-photon helicities respectively.
Energy distribution in terms of the variable $y$ is related to that in
terms of $\omega$, for fixed $E_b$ and $\omega_0$, by
\bea
f(\omega) = \f{1}{\sg_c}\,\f{d\sg_c}{d\omega}=\f{1}{E_b}\:f(x,y).
\label{eq:edist}
\eea
The total cross section $\sigma_c$ is given by
\[\sigma_c=\sigma_c^{np}+2\lae \ll \,\sg_1, \]
with
\[
\sg_c^{np}=\f{2 \pi \alpha^2}{x m_e^2}\left[\left(1-\f{4}{x}-
\f{8}{x^2}\right)\log (x+1) +\f{1}{2}+\f{8}{x}-\f{1}{2(x+1)^2}\right]
\]
and
\[
\sg_1=\f{2 \pi \alpha^2}{x m_e^2}\left[\left(1+\f{2}{x}\right)
\log (x+1)-\f{5}{2}+ \f{1}{1+x}-\f{1}{2(x+1)^2}\right].
\]
Here $\sg_c^{np}$ is the unpolarized cross section.
The energy spectrum $f(x,y)$ plotted against $y$ is shown in Figure
\ref{fig:dist_pol}.

It is clear from the figure that when $\lae \ll < 0$,
there are more number of hard photons than soft photons, while 
for $\lae \ll > 0$ the number of hard photons is less than the 
number of soft photons. Also in the case of $\lae \ll < 0$
the spectrum peaks at higher energies, resulting in nearly 
monochromatic beams.

Polarized photon beams are more appropriate for \cp\ violation studies.
Dependence of the helicity of the Compton-scattered photon on the energy of
the photon is discussed in \cite{ginzburg} and is given by 
\bea
\lg (\om)=\f{\ll\:(1-2 r)\:(1-y+\f{1}{1-y})+2\lae\:r\:x\:\left[
1+(1-y)\:(1-2\:r)^2\right]}
{1-y+\f{1}{1-y}-4\,r\:(1-r)-2\lae\:\ll\:r\:x\:(2\:r-1)\:(2-y)}.
\label{eq:lambda}
\eea
For $2\,\lae\ll=-1,$ hard photons will have helicity $\lambda=\lae$ (See Figure 
\ref{fig:dist_pol}). As already mentioned the number of hard photons is 
much higher than that of the soft ones at $2\,\lae \ll =-1$.

\begin{figure}
\vskip 6cm
\includegraphics{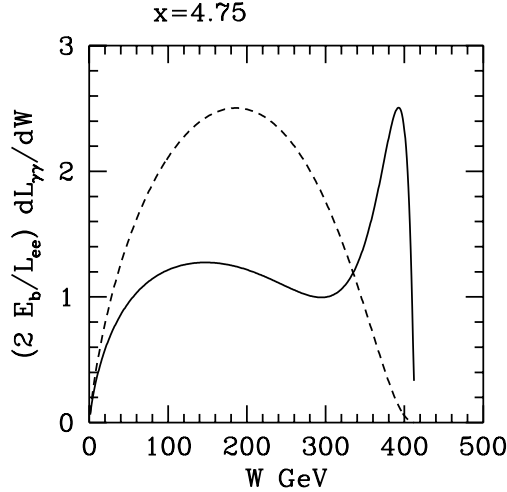}
\caption[dummy]
{\small Luminosity distributions are plotted against the $\gg$ invariant mass,
\(
W=2\sqrt{\om_1 \om_2}
\). The initial electron beam energy, $E_b$ is taken to be 250 GeV and a laser
beam of energy 1.24 eV is assumed. The solid curve is for $2\,\lae \ll=-1$ while
the dotted curve is for $2\,\lae \ll=1$.  The conversion distance is assumed to 
be zero.}
\label{fig:lumdist} 
\end{figure}

Another important aspect of a collider is its luminosity.
The luminosity distribution of a $\gg$ collider depends on different factors
like the conversion distance, {\em i.e.}, the distance from the 
scattering point to the interaction point, energy distribution of 
the beams, etc. Assuming a Gaussian profile for the electron beam with
azimuthal symmetry, the luminosity distribution of a $\g \g$ collider is 
given in terms of the photon energy distribution by \cite{ginzburg}
\bea
\f{1}{L_{ee}}\,\f{dL_{\g \g}}{d\om_1\,d\om_2}=
	f_1(\om_1)\,f_2(\om_2)\;
	I_0\left( \f{d_1d_2}{\sg_1^2+\sg_2^2}\right)\;
	e^{-\f{d_1d_2}{2(\sg_1^2+\sg_2^2)}}.
\label{eq:lum1}
\eea
Here $f_1(\om_1)$ and $f_2(\om_2)$ are the energy distributions of the two
photon beams (see eqs.~(\ref{eq:endist}) and (\ref{eq:edist})).
$I_0$ is the zeroth order modified Bessel function with
$d_i=z_i\theta_{\g i}$, where $z_i$ is the conversion distance and
$\theta_{\g i} $ is the scattering angle of the photon beam,
and $\sg_i$ is the half width of the Gaussian profile.
$L_{ee}$ is the geometric luminosity of the original electron-electron
collider. 
Making a variable change from $\om_1$ and $\om_2$ to $\eta$ and $W$,
where
\( \eta=tan^{-1}\left(\f{\om_1-\om_2}{\om_1+\om_2}\right) 
\)
is the $\gg$ rapidity and 
\(
W=2\sqrt{\om_1 \om_2}
\)
is the $\gg$ invariant mass, we get the luminosity distribution as 
\bea
\f{1}{L_{ee}}\,\f{dL_{\g \g}}{dW\,d\eta}=
	\f{W}{2}\;f_1\left(\f{We^{\eta}}{2}\right)\,
	f_2\left(\f{We^{-\eta}}{2}\right)\;
	I_0\left( \f{d_1d_2}{\sg_1^2+\sg_2^2}\right)\;
	e^{-\f{d_1d_2}{2(\sg_1^2+\sg_2^2)}}.
\label{eq:lum2}
\eea
Taking the conversion distance to be zero for simplicity, the expression 
for the luminosity distribution becomes
\bea
\f{1}{L_{ee}} \f{dL_{\g \g}}{dW d\eta}=
        \f{W}{2}f_1\left(\f{We^{\eta}}{2}\right)
	f_2\left(\f{We^{-\eta}}{2}\right).
\label{eq:lum}
\eea
Figure~\ref{fig:lumdist} gives the luminosity distribution after rapidity is 
integrated out.
The luminosity peaks at higher values of invariant mass in case of $2\,\lae \ll
=-1$ and the peak value could be as high as 90\% of $L_{ee}$,
while for $2\,\lae \ll=1$ the spectrum is almost a Gaussian  peaking at
low energies.

The expression for the number of events in a particular process $\g \g \ra X$
for the general case of arbitrary electron and laser-photon polarizations is
quite complicated. Considerable simplification results with the following
assumptions: (i) Axial symmetry of the beam (ii) Only longitudinal polarization
of the electron beams (iii) Only circular polarization of the laser beams (iv)
Negligible distance between the conversion points and the interaction points.
In that case, the converted photons have only circular polarizations given by
the average values of the Stokes parameter $\xi_2$. The number of events is 
then given in terms of the
luminosity distribution (eq.~(\ref{eq:lum})) and the average value of the 
Stokes
parameters $\xi_2$ and $\o{\xi}_2$ of the two photon beams, by
\cite{ginzburg2} 
\bea
dN_{\gg \ra X}&=&dL_{\gg}\;\left(d\sg_{00}+\xi_2\o{\xi}_2d\sg_{22}
       +\xi_2d\sg_{20}+\o{\xi}_2d\sg_{02}\right).
\label{eq:dsij}
\eea
Expressions for $d\sg_{ij}$, which are linear combinations of production
density-matrix elements,  and $\xi_2$ and $\o{\xi}_2$ are given in the 
appendix.

For large values of $L_{ee}$, which are possible to achieve, we expect 
large \ttbar\ production in a $\gg$ collider.  In the next 
section we shall discuss some of the \cp-violating effects which could 
be tested in these colliders.

\section{Charge Asymmetries in  \ggtt}

We consider the effective \l\ 
\bea
\lc_{\it eff}&=&\lc_{SM}+\lc_{CP},
\label{eq:efflag}
\eea
where $\lc_{SM}$ is the usual SM Lagrangian, and 
\bea
\lc_{CP}&=&ie\,d_t\,\o{\psi}_t\,\sigma^{\mu \nu}\,
\gamma_{\small 5}\,\psi_t\;F_{\mu \nu} \label{eq:lcp}
\eea
with
\bea
F_{\mu \nu} = \partial_{\mu}\,A_{\nu}-
\partial_{\nu}\,A_{\mu}.
\eea
$d_t$ is the electric dipole
form factor and is, in general, complex and momentum dependent.
This modifies the SM $t \o t \g$ coupling to
$ie\Gamma_{\mu}$, where
\bea
\Gamma_\mu\;=\;\f{2}{3}\,\gamma_\mu\;+
\;d_t\,\sigma_{\mu \nu}\,\gamma_{\small 5}\,(p_t\,+\,p_{\bar{t}})^{\nu}.
\label{eq:effvertex}
\eea
\cp\ violation arising due to this electric dipole moment of the top quark
could be studied by using \cp-violating asymmetries in the processes
involving top-quark coupling with photons. We consider the process \ggtt\
with the subsequent decay of $t$ and $\o{t}$. We neglect $CP$ violation in the
decay of $t$ and $\o{t}$.

The asymmetries which do not depend on the top quark momentum that we 
consider here
had been considered by us earlier in the context of an \ep\ collider 
\cite{paul2}.
In simple terms these asymmetries are (i) the asymmetry in the number of
leptons and antileptons produced as decay products of top antiquark and top
quark (the charge asymmetry) and (ii) the sum of the forward-backward
asymmetries of the leptons and antileptons. These asymmetries being 
independent of the top-quark momentum are experimentally favourable.
The charge asymmetry is zero in the absence of a cut-off in the polar angle
of the lepton (antilepton). This is because when the cut off is zero the
charge asymmetry is just the asymmetry in the production rates of $t$ and
\tbar\, which is zero from charge conservation assuming that $CP$ is not
violated in $t$ decay.

The two asymmetries are written in terms of differential cross section as
follows.
\be
A_{ch}(\theta_0)=\frac{
{\displaystyle          \int_{\theta_0}^{\pi-\theta_0}}d\theta_l
{\displaystyle          \left( \frac{d\sigma^+}{d\theta_l}
                -       \frac{d\sigma^-}{d\theta_l}\right)}}
{
{\displaystyle          \int_{\theta_0}^{\pi-\theta_0}}d\theta_l
{\displaystyle          \left( \frac{d\sigma^+}{d\theta_l} +
\frac{d\sigma^-}{d\theta_l}\right)}}
\label{eq:ach}
\ee
and
\be
A_{fb}(\theta_0)= \frac{ {\displaystyle
\int_{\theta_0}^{\frac{\pi}{2}}}d\theta_l {\displaystyle
\left( \frac{d\sigma^+}{d\theta_l} +
\frac{d\sigma^-}{d\theta_l}\right)} {\displaystyle
-\int^{\pi-\theta_0}_{\frac{\pi}{2}}}d\theta_l {\displaystyle
\left( \frac{d\sigma^+}{d\theta_l} +    \frac{d\sigma^-}{d\theta_l}
\right)}}
{
{\displaystyle          \int_{\theta_0}^{\pi-\theta_0}}d\theta_l
{\displaystyle          \left( \frac{d\sigma^+}{d\theta_l} +
\frac{d\sigma^-}{d\theta_l}\right)}}.
\label{eq:achfb}
\ee
In the above equations, $\frac{d\sigma^+}{d\theta_l}$ and 
$\frac{d\sigma^-}{d\theta_l}$ refer respectively to
the $l^+$ and $l^-$ distributions in the c.m. frame
and $\theta_0$ is the cut-off.

Both the asymmetries defined above are even under naive time reversal $T_N$, 
which changes the signs of spins and momenta, without interchange of the 
initial and final states.
They are therefore odd under the combination $CPT_N$. To avoid conflict with
the $CPT$ theorem, the asymmetries can only depend on the imaginary part of
$d_t$.

We consider only the semileptonic decay of \ttbar. This means that either of $t$
or \tbar\ decays into a $b$ or $\overline{b}$ and leptons,
while the other decays
hadronically. We choose this semi-leptonic mode for two reasons. 
Firstly, the leptonic decay (into $\mu$ or $e$) is cleaner to trigger on,
and secondly, allowing the other decay to be hadronic gives a larger number of
events due to the larger branching ratio. 
We work in the narrow-width approximation, where
$W$ boson is produced on-shell. We also consider on-shell production of
$t$ and \tbar\ which allows us to separate the production and decay parts of the
amplitude. 

\begin{figure}
\vskip 45mm
\includegraphics{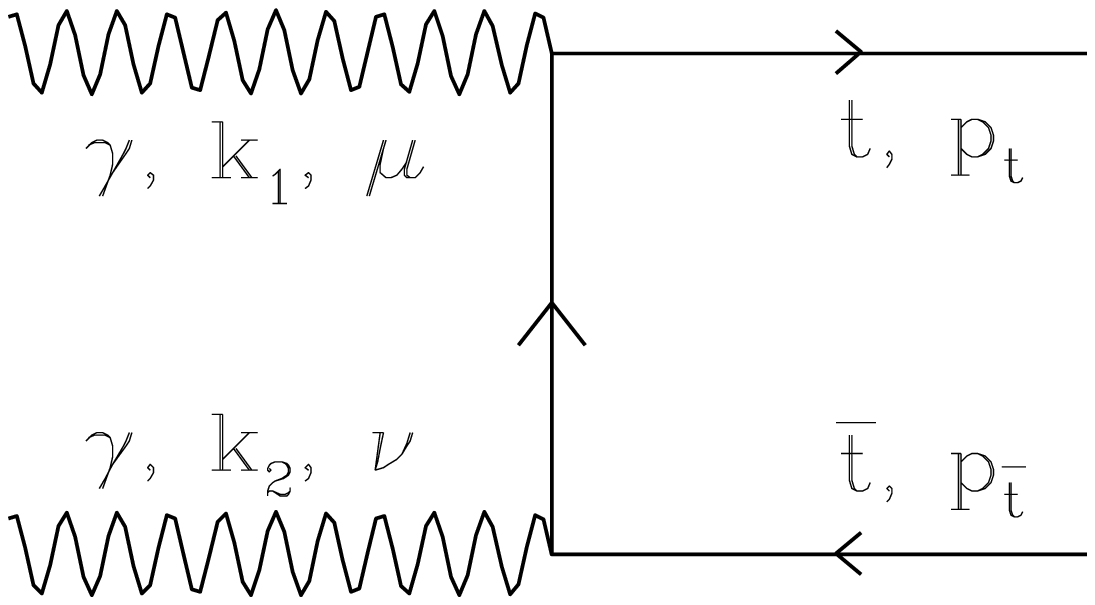}
\includegraphics{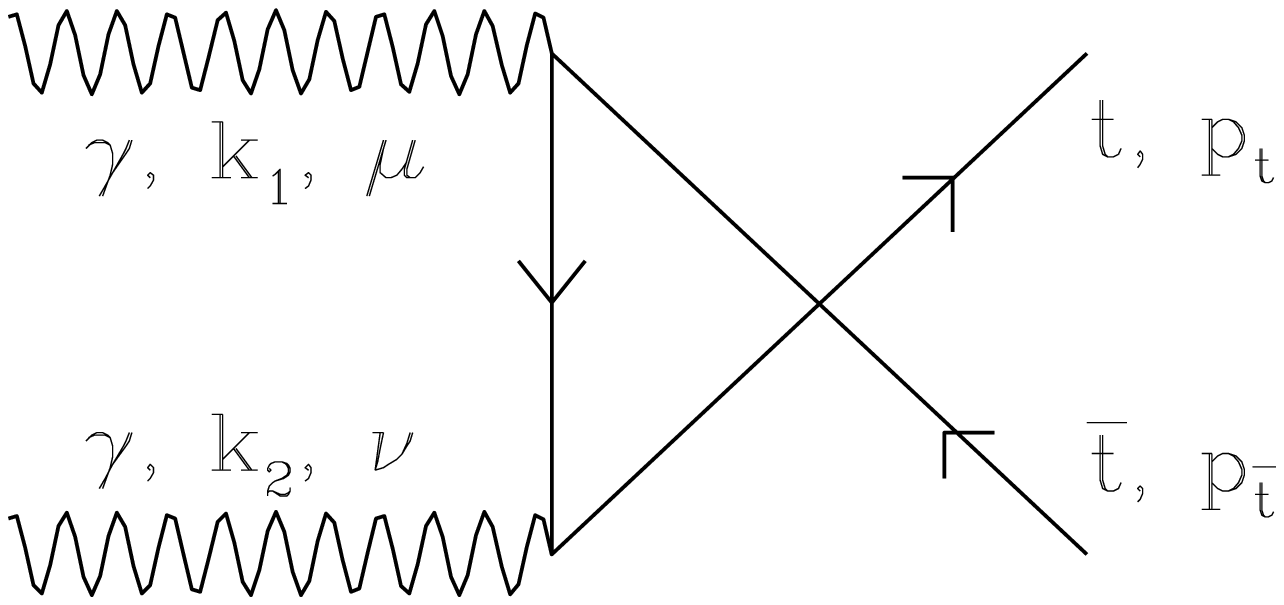}
\caption[dummy]
{Feynman diagrams for the process \ggtt\.}
\label{fig:feyndiag5}
\end{figure}

The Feynman diagrams for the process
$\gamma_{\mu}(k1)\gamma_{\nu}(k2) \rightarrow 
t(p_t)\overline{t} (p_{\overline{t}})$ 
are shown in Figure
\ref{fig:feyndiag5}.

We make us of helicity amplitudes. Thus, our approach is different 
from that of \cite{tsai}, which uses trace techniques, and which was made use 
of, for example, in \cite{arens} in the case of $e^+e^-\rightarrow t\o{t}$.
The production helicity amplitudes are calculated using a
method developed by Vega and Wudka \cite{wudka}. The helicity amplitudes
are given by
\bea
M(\lg,~\lg,\lt,~\lt)&=& -\f{4 m_t\,e^2\,Q_t^2}{\sqrt{s}
	(1-\bt^2\cos^2\theta_t)}
        \left\{(\lg+\lt \bt) \right. \non \\
        &&-i\,d_t\;2m_t\left[2+\f{s}{4m_t^2}\bt
        (\bt-\lt \lg) \sin^2\theta_t\right] \non \\
        &&+d_t^2\;\f{s\lg}{2}\,\left. \left[\f{4 m_t^2}{s}+\bt(\bt-\lg \lt)
       \sin^2\theta_t\right]\right\}\non \\
M(\lg,~\lg,\lt,-\lt)&=& -\f{4
	m_t\,e^2\,Q_t^2}{(1-\bt^2\cos^2\theta_t)} \non \\
        &&\times \bt
        \sin\theta_t\;\cos\theta_t\;\left[ \lg\,i\,d_t-m_t\,d_t^2
	\right]\non \\
M(\lg,-\lg,\lt,~\lt)&=& \f{4 m_t\,e^2\,Q_t^2}{\sqrt{s}
	(1-\bt^2\cos^2\theta_t)} \non \\
        &&\times \left[ \lt\,\bt+i\,d_t\;\f{s}{2m_t}\bt^2-
        d_t^2\;\f{s}{2}\lt \bt \right]\;\sin^2\theta_t \non \\
M(\lg,-\lg,\lt,-\lt)&=&
	\f{2\bt\,e^2\,Q_t^2}{(1-\bt^2\cos^2\theta_t)}\sin\theta_t
        \left\{\left( \lg \lt +\cos\theta_t\right.\right) \non \\
        &&-d_t^2\;\f{s}{2}\left. \left[\f{4m_t^2}{s}\,\cos\tht+
        \lg \lt (1-\bt^2\cos^2\theta_t) \right]\right\}.
\label{eq:ggttamp}
\eea
The notation is that in 
$M({\lg}_1,{\lg}_2,\lt,\ltb)$, ${\lg}_1,\,\,{\lg}_2,\,\,\lt$ and 
$\ltb$ correspond to the helicities of the two incoming photons, the top quark 
and the
top antiquark respectively. $Q_t=2/3$ is the top charge, $\theta_t$ 
is the scattering angle in the c.m.
frame, and $\bt$ is the top-quark velocity.
It should be noted that $d_t$ occurring in eq.~(\ref{eq:ggttamp}) 
is in reality a dipole form factor and not the on-shell 
dipole moment, because in Figure \ref{fig:feyndiag5}, one top line at each
vertex is off-shell. 
These expressions agree with those in \cite{Baek}.

To be able to calculate from these amplitudes the differential cross section 
of the complete process with
initial electron states using eq. (\ref{eq:dsij}), we first need
to know the combinations of cross sections $d\sg_{ij}$ for the 
$\gamma\gamma$ subprocess.
These we write in terms of the
production density matrix elements $\rho_{ij}$, expressions for which are
given in the appendix. We get the following expressions.
\bea
\frac{d\sigma_{ij}^{\pm}}{d\cos\theta_t\, dE_l\, d\cos\theta_l\,
d\phi_l}&=& \frac{3\alpha^2\beta}{16x_w^2\sqrt{s}}\,
 \frac{E_l}{\Gamma_t \Gamma_W m_W}\,
\left(\frac{1}{1-\beta\cos\theta_{tl}}
  -\frac{4 E_l}{\sqrt{s}(1-\beta^2)}\right)\nonumber \\
&&\times \left\{
\left[ {\rho}_{ij}^\pm(++)+{\rho}_{ij}^\pm(--)\right]
(1-\beta \cos\theta_{tl})\right. \nonumber \\
&&\left. +
\left[ {\rho}_{ij}^\pm(++)-{\rho}_{ij}^\pm(--)\right]
(\cos\theta_{tl}-\beta )\right. \nonumber
\\
&&+ 2\,{\rm Re}\,\left({\rho}_{ij}^\pm\left(+-\right)\right)\,
(1-\beta^2) \non\\
&&\times \sin\theta_t \sin\theta_l \left(\cos\theta_t \cos\phi_l -
\sin\theta_t \cot\theta_l\right) \non\\
&&\left. + 2\,{\rm Im}\,\left({\rho}_{ij}^\pm\left(+-\right)\right)\,
(1-\beta^2) \sin\theta_t \sin\theta_l \sin\phi_l \right\}.
\label{eq:csectiongg}
\eea
Here $\theta_{tl}$ is the angle between the top quark and the lepton, and
$\theta_l$ and $\phi_l$ are the polar and azimuthal angles of the lepton
in the c.m. frame.
We consider only terms at most linear in $d_t$. All the higher-order terms
are neglected assuming that $d_t$ is small.
The superscript $\pm$ correspond respectively to $t$ and \tbar\ decaying
into leptons.

The cross section is given in terms of $d\sg_{ij}^{\pm}$ as 
\bea
d\sg^{\pm}&=&d\sg^{\pm}_{00}+\xi_2\o{\xi}_2d\sg^{\pm}_{22}
        +\xi_2d\sg^{\pm}_{20}+\o{\xi}_2d\sg^{\pm}_{02}.
\eea

We now go over to $CP$-odd asymmetries which can be obtained from the
differential cross section.
The charge asymmetry is defined as
\bea
A_{ch}&=&\f{1}{2\,N}\left\{\int \f{d L_{\g \g}}{d\om_1
d\om_2}\,d\om_1\,d\om_2
        \int_{-1}^1\!d\cos\tht\right. \non \\
        &&\times \left.
        \int_{\theta_0}^{\pi-\theta_0}
        d\theta_l\left[ \frac{d\sigma^-} {d\cos\tht\,d\theta_l}(\theta_l)-
     \frac{d\sigma^+}{d\cos\tht\,d\theta_l}(\pi-\tl)\right]\right\}.
\label{eq:ggach1}
\eea
The charge asymmetry combined with the forward-backward asymmetry is
defined as
\bea
A_{fb}&=&\f{1}{2\,N}\int \f{d L_{\g \g}}{d\om_1
d\om_2}\,d\om_1\,d\om_2
        \int_{-1}^1\!d\cos\tht \non \\
        &&\times \left\{
        \int_{\theta_0}^{\pi/2}
        d\theta_l\left[ \frac{d\sigma^-} {d\cos\tht\,d\theta_l}(\theta_l)+
    \frac{d\sigma^+}{d\cos\tht\,d\theta_l}(\pi-\tl)\right]\right.\non \\
	&&-\left.\int_{\pi/2}^{\pi-\theta_0}
        d\theta_l\left[ \frac{d\sigma^-} {d\cos\tht\,d\theta_l}(\theta_l)+
    \frac{d\sigma^+}{d\cos\tht\,d\theta_l}(\pi-\tl)\right]\right\}
\label{eq:ggafb1}
\eea
$N$ is the total number of events given by integrating $dN$ in eq.
(\ref{eq:dsij}) with $d\sg_{ij}$ given by eq.~(\ref{eq:csectiongg}).
$\omega_1$ and $\omega_2$ are the energies of the two photon beams.

The expression for the angular distribution given by 
eq.~(\ref{eq:csectiongg}) is in the $\g \g$ c.m. frame whereas the 
expressions
for \ach\ and $A_{fb}$ above (eqs.~(\ref{eq:ggach1}) and (\ref{eq:ggafb1})) are
in the lab frame. We will therefore rewrite these latter expressions 
in the $\g\g$ c.m. frame.
Changing the variable of integration to $\cos\tl$ and noticing that the
lab frame is obtained by boosting the c.m. frame by a velocity 
$\bg=\f{\om_1-\om_2}
{\om_1+\om_2}$, we get the lower and the upper limits of integration in the
$\g\g$ c.m. frame as
\[
f(\theta_0)=\f{\cos\theta^{cm}_0+\bg}{1+\bg \cos\theta^{cm}_0}
\] and
\bea
g(\theta_0)&=&\f{\cos(\pi-\theta^{cm}_0)+\bg}{1+\bg
        \cos(\pi-\theta^{cm}_0)} \non \\
&=&\f{-\cos\theta^{cm}_0+\bg}{1-\bg \cos\theta^{cm}_0}. \non
\eea
Making use of the fact that
\[
f(\pi-\theta_0)=\f{-\cos\theta^{cm}_0+\bg}
        {1-\bg\,\cos\theta^{cm}_0} =g(\theta_0)\non
\]
and
\[
g(\pi-\theta_0)=\f{\cos\theta^{cm}_0+\bg}{1+\bg\, \cos\theta^{cm}_0}
=f(\theta_0),
\]
we get the final expression for \ach\ as
\bea
A_{ch}&=&\f{1}{2\,N}\left\{\int \f{d L_{\g \g}}{d\om_1\,d\om_2}\,d\om_1\,d\om_2 
	\int_{-1}^1\!d\cos\tht\right.\non \\
        &&\times\left.
	\int_{f(\theta_0)}^{g(\theta_0)}
	d\cos\tl\left[\frac{d\sigma^-}{d\cos\tht\,d\cos\tl}\,(\tl)-
      \frac{d\sigma^+}{d\cos\tht\,d\cos\tl}(\tl)\right]\right\},
\label{eq:ggach2} 
\eea
where the differential cross section is in the c.m. frame.
A similar expression holds for \afb :
\bea
A_{fb}&=&\f{1}{2\,N}\,\int \f{d L_{\g \g}}{d\om_1\,d\om_2}\,d\om_1\,d\om_2 
	\int_{-1}^1\!d\cos\tht\non \\
        &&\times \left\{
	\int_{f(\theta_0)}^{\beta_\g}
	d\cos\tl\left[ \frac{d\sigma^-}{d\cos\tht\,d\cos\tl}\,(\tl)+
      \frac{d\sigma^+}{d\cos\tht\,d\cos\tl}\,(\tl)\right]\right.\non \\
	&&\left.-
	\int_{\beta_\g}^{g(\theta_0)}
	d\cos\tl\left[ \frac{d\sigma^-}{d\cos\tht\,d\cos\tl}\,(\theta_l)+
      \frac{d\sigma^+}{d\cos\tht\,d\cos\tl}\,(\tl)\right]\right\}.
\label{eq:ggafb} 
\eea

We use the expressions derived above to evaluate the sensitivity of the
experimental set-ups with different luminosities and energies to the top dipole
moments.
The sensitivity of an experiment to the measurement of a physical quantity 
depends on the statistics. The number of
asymmetric events must be greater than the statistical fluctuation by a
certain factor for it to be observed. This factor determines the confidence
level (C.L.) of the measurement.  For a system with one degree of freedom
the number of asymmetric events $N_A\;(=N\,A,$ where $A$ is the
asymmetry and $N$ is the total number of events) for a certain value of dipole 
moment $d_t$ should then be
greater than $1.64 \sqrt{N}$ for the dipole moment to be observable 
at the 90\% C.L., where
$\sqrt{N}$ corresponds to the standard deviation.
Using this we can get limits that can be set on the dipole form factors in case
the asymmetry is not observed.
The 90\% C.L. on the dipole form factor is therefore given by
\bea
\delta d_t = \f{1.64 d_t}{\sqrt{N}A} = \f{1.64}{\sqrt{N}{\cal A}},
\label{eq:sensitivity1}
\eea
where ${\cal A}=A/d_t$ is the value of the asymmetry $A$ for unit
dipole moment.

The following section discusses the results we obtain. We have carried out the
$\cos\theta_l$ integrals in the above expressions analytically, and the
remaining integrals, viz., those over $\cos\theta_t$, $\omega_1$ and
$\omega_2$, numerically.

\section{Results and Conclusion}

Charge asymmetry and forward-backward asymmetry combined with
charge asymmetry are calculated for different initial-beam helicities. Also,
fixing a particular helicity combination, asymmetries are obtained at 
different electron beam energies and for different laser beam energies. In 
doing so, the value of $x$ is kept constant. Variation of asymmetries with $x$,
fixing variables like the helicites and cut-off angle is also studied.
Asymmetries are also studied at different cut-off angles with beam energy
and other parameters kept constant. All the calculations are done 
assuming a geometrical integrated luminosity of 20 fb$^{-1}$ for the 
electron-electron collider.  We also assume the laser energy $\omega_0$ 
to be 1.24 eV. We discuss the results in the following.

Table \ref{tab:polcomb} displays asymmetries 
(in all cases we calculate asymmetries for a fixed value of $\idt =
1/(2m_t)$) obtained for different 
helicity combinations for beam energy 250 GeV and cut-off angle $30^{\small
\circ}$. The top-quark mass is taken to be 174 GeV. Some important features of
the results in Table \ref{tab:polcomb} may be noted. 
There is no combined asymmetry when both $\lambda_e^1={\lae}^2$ and 
$\lambda_l^1=\lambda_l^2$.  This is expected as the forward and backward 
directions cannot be 
distinguished in this case because the two colliding photons are
identical. In SM electromagnetic interactions respect parity and hence 
the cross section is symmetric under ${\lae}^i\leftrightarrow 
-{\lae}^i$. Thus the total number of events, which gets contribution only
from SM, remains the same under this transformation.

\begin{table}
\begin{center}
\begin{tabular}{rrrrrcccc}
\hline
\\
&&&&&
\multicolumn{2}{c}{Asymmetries}&
\multicolumn{2}{c}{Limits on Im $d_t$ }\\
$\lambda_e^1$&$\lambda_e^2$&$\lambda_l^1$&$\lambda_l^2$&$N$&
$A_{ch}$& $A_{fb}$&
\multicolumn{2}{c}{($10^{-16}\;e\,cm$) from}\\
&&&&&&&$|A_{ch}|$&$|A_{fb}|$\\[2mm]
\hline
\\
$-.5$&$-.5$&$-1$&$-1$& 76& $-.019$ &0 &  2.76 & \\
$-.5$&$-.5$&1&$-1$& 252& $ -.025$& $ -.129$&  1.19&  .230\\
$-.5$&$-.5$&1& 1& 631& $ -.035$& 0&  .54 & \\
.5&$-.5$&$-1$&$ -1$& 73&$ -.024$ &  .013&  2.31&   4.25\\
.5&$-.5$& 1&$ -1$& 32&$  -.021$ & $-.080$ & 3.89& 1.03\\
.5&$-.5$&$-1$&  1& 163& $ -.021$&   .033&  1.73 &  1.12\\[3mm]
\multicolumn{4}{c}{Unpolarized}&194&$ -.028$&0& 1.19&\\[3mm]
\hline
\end{tabular}
\end{center}
\caption{
Asymmetries and corresponding 90\% C.L. limits obtained on Im$\,d_t$ for 
various combinations of initial beam helicities. The top mass is kept at 
$m_t = 174$ GeV and an initial electron beam of energy
$E_b  =  250$ GeV and a laser beam of energy $\omega_0 = 1.24$ eV are 
assumed.
The cut-off angle taken is $\theta_0=30^\circ$. $N$ is the total number of
events.  Asymmetries are for ${\rm Im}\,d_t=\f{1}{2m_t}$.
}
\label{tab:polcomb}
\end{table}

As seen from Table \ref{tab:polcomb}, the 90\% C. L. limits on Im$\,d_t$
are in general of the order of $10^{-16}\;e\:cm$.
The limit obtained on Im$\,d_t$ in the unpolarized case is $1.19\times
10^{-16}\;e\:cm$.  Forward-backward combined asymmetry is zero 
in this case.
The best limit obtained is however an order of magnitude better. Its value is 
$2.3\times10^{-17}\;e\:cm$ and comes from
\afb\ with initial-beam helicities satisfying ${\lae}^1={\lae}^2$ and
$\lambda_l^1=-\lambda_l^2$. 
We consider this helicity combination for 
further analysis.  It may be noted, however, that with unpolarized electron 
beams and laser beams we can measure only the charge asymmetry. 

Table~\ref{tab:en} lists the limits for different electron beam energies.
The table shows that the limits
are better around a beam energy of 500 GeV in the case of combined
asymmetry and at around 750 GeV in the case of charge asymmetry. 
The limit obtained at this
value is almost 20 times better than the limit at 250 GeV in the 
case of charge asymmetry while in the case of combined asymmetry it is 
a factor of almost 8. 

\begin{table}
\begin{center}
\begin{tabular}{rrrrrr}
\hline
&&&\\
&&
\multicolumn{2}{c}{Asymmetries}&
\multicolumn{2}{c}{Limits on Im $d_t$}\\
\multicolumn{1}{c}{$E_b$}&\multicolumn{1}{c}{$N$}&$A_{ch}$&$A_{fb}$&
\multicolumn{2}{c}{($10^{-16}\;e\,cm)$ from}\\
(GeV)&&&&$|A_{ch}|$&$|A_{fb}|$\\[3mm]
\hline
&&&\\
250 &252 & --.025& .129& 1.19& .230\\
500 &1441& --.167& .420& .074 & .029\\
750 &1210& --.223& .347& .060 & .039\\
1000&996 & --.227& .244& .065 & .061\\[3mm]
\hline
\end{tabular}
\end{center}
\caption{
Variation of limits on Im$\,d_t$ obtained at different beam energies keeping $x$
fixed at 4.75 (by choosing suitable laser beam energy in each case).
The top mass used is 174 GeV and the cut-off angle is taken to
be $30^\circ$. Asymmetries are for ${\rm Im}\,d_t=\f{1}{2m_t}$. 
Helicities of the initial electron and laser beams
are $\lambda_e^1=-.5,\;\lambda_e^2=-.5,\;\lambda_l^1=-1$ and 
$\lambda_l^2= 1$.}
\label{tab:en}
\end{table}

The cut-off angle is also varied to study the variation of limits on 
Im$\,d_t$. The
result is tabulated in Table~\ref{tab:ang}. As mentioned before, charge 
asymmetry, which is the total 
leptonic charge in 
the semi-leptonic decay of \ttbar\ is zero when there is no cut-off.
Charge asymmetry is found to give best limits on the dipole form factors 
around  a cut-off of $60^\circ$ whereas the combined asymmetry is better at
lower cut-offs. 
\begin{table}
\begin{center}
\begin{tabular}{rrrrrr}
\hline
&&&\\
&&
\multicolumn{2}{c}{Asymmetries}&
\multicolumn{2}{c}{Limits on Im $d_t$}\\
\multicolumn{1}{c}{$\theta_0$}&\multicolumn{1}{c}{$N$}&$A_{ch}$&$A_{fb}$&
\multicolumn{2}{c}{($10^{-16}\;e\,cm)$ from}\\
(deg.)&&&&$|A_{ch}|$&$|A_{fb}|$\\[3mm]
\hline
&\\
    0& 290&  .000& .149&       & .185\\
   10& 286& --.003& .146& 9.23 & .189\\
   20& 273& --.012& .140& 2.44 & .203\\
   30& 252& --.025& .128& 1.19 & .230\\
   40& 221& --.041& .113& .774  & .277\\
   50& 186& --.057& .095& .599  & .362\\
   60& 144& --.073& .074& .534  & .529\\
   70&  98& --.086& .050& .551  & .937\\
   80&  50& --.094& .026& .706  & 2.595\\[3mm]
\hline
\end{tabular}
\end{center}
\caption{Limits on Im$\,d_t$ of the top quark from the charge asymmetry and
the combined asymmetry for different cut-off angles. 
Helicities of the initial electron and laser beams
are $\lambda_e^1=-.5,\;\lambda_e^2=-.5\;\lambda_l^1=-1$ and $
\lambda_l^2= 1$. A top-quark mass of 174 GeV and an electron beam energy of
250 GeV are used. Laser beam energy is taken to be 1.24 eV, which
corresponds to $x=4.75$. Asymmetries are for ${\rm Im}\,d_t=\f{1}{2m_t}$.
}
\label{tab:ang}
\end{table}

For a fixed beam energy and at a fixed cut-off angle, variation of Im$\,d_t$
limits with $x$ is studied in Table~\ref{tab:xvalue}. From the table it is 
clear that
the limits are better at higher $x$ values. In fact, for $x$ less than about 3,
the event rate is quite low, and no significant limit on the dipole moment may
be expected. On the other hand, for $x>4.83$, \ep\
production due to the collision of high energy
photon beam with laser beam is considerable \cite{ginzburg}.
This introduces additional \ep\ beam backgrounds as well as degrading the
photon spectrum.

\begin{table}
\begin{center}
\begin{tabular}{rrrrrr}
\hline
&&&\\
&&
\multicolumn{2}{c}{Asymmetries}&
\multicolumn{2}{c}{Limits on Im $d_t$}\\
\multicolumn{1}{c}{$x$}&\multicolumn{1}{c}{$N$}&$A_{ch}$&$A_{fb}$&
\multicolumn{2}{c}{($10^{-16}\;e\,cm)$ from}\\
&&&&$|A_{ch}|$&$|A_{fb}|$\\[3mm]
\hline
&\\
2.60&  1.1& --.004 & .034& 113 &12.8\\
3.20& 28.8& --.011 & .072&   7.85 & 1.22\\
4.74&250.8& --.025 & .128 &   1.20 &  .230\\[3mm]
\hline
\end{tabular}
\end{center}
\caption{
Limits on Im$\,d_t$ calculated at different $x$ values for a fixed beam energy
$E_b  =  250$ GeV and for a cut-off angle of $30^\circ$. The top mass is
taken to be  174 GeV and the helicities of the initial beams are 
${\lae}^1=-.5,\;{\lae}_\gamma^2=-.5,\,\lambda_l^1=-1$ and
$\lambda_l^2=1$.}
\label{tab:xvalue}
\end{table}

To conclude, we find that for an electron beam energy of 250 GeV, and for a
suitable choice of circular polarizations of the laser photons and longitudinal
polarizations for the 
electron beams, and assuming a geometrical luminosity of 20 fb$^{-1}$
for the electron beam, it is possible to obtain limits on the imaginary
part of the top EDFF of the order of $10^{-17}\;e\,cm$.  An order of
magnitude improvement is possible if the beam energy is increased to 500
GeV.  In that case the sensitivity would be comparable to that obtained in
\eett\ with $\sqrt{s}=500$ GeV \cite{gzad,paul1,paul2} .

For comparison, we mention the result of Baek {\it et al.} \cite{Baek} 
(who have extended
the work of \cite{choigg}, and corrected numerical errors therein). 
They have obtained limits on the real part
of the dipole moments of the top quark from a number asymmetry and with
linearly polarized photons to be of the order of $10^{-17}\;e\,cm$
for a beam energy of 250 GeV.

The sensitivities discussed above are under ideal experimental conditions, 
which may not be met in practice. A more detailed calculation taking into
account experimental cuts and realistic choices of parameters like laser-photon
energies, degrees of polarization of the beams, etc. would be important to
carry out. However, the results obtained here would be indicative of the
possibilities in a realistic case.

The asymmetries we discuss only give limits on the imaginary part of the top
EDFF. It would be interesting to study other asymmetries which give limits on 
the real part of $d_t$ in the presence of circularly polarized photon beams.
\vskip 1cm
\ni
{\Large  \bf Acknowledgement} 

\vskip .2cm
We thank V. Ravindran for collaboration in
the initial stages of the work.

\vskip 15mm
\ni
{\LARGE  \bf Appendix}
\appendix

\section{Expressions for $d\sg_{ij}$ and $\xi_2$ for $\gamma\gamma\rightarrow 
X$}
\label{sec:dsij}

Expressions for $d\sg_{ij}$ and $\xi_2$ 
for a general process $\gamma\gamma
\rightarrow X$, referred to in eq.~(\ref{eq:dsij}),
are given here in terms of the 
amplitudes. 

$d\sigma_{ij}$ are given by
\bea
d\sg_{00}&=&\f{1}{4}\,\sum_{\lg,\lg'}|M(\lg,\lg')|^2 \,d\Gamma\non \\
d\sg_{22}&=&\f{1}{4}\,\sum_{\lg,\lg'}\lg \lg'\,|M(\lg,\lg')|^2 \,d\Gamma\non \\
(d\sg_{20}+d\sg_{02})&=&\f{1}{2}\,\sum_{\lg}\lg \,|M(\lg,\lg)|^2 \,d\Gamma
\non \\
(d\sg_{20}-d\sg_{02})&=&\f{1}{2}\,\sum_{\lg}\lg \,|M(\lg,-\lg)|^2 \,d\Gamma
\non 
\eea
Helicities of the particles in the final state are summed over, and
not shown. $d\Gamma$ is the appropriate phase space factor for the final state
$X$.

The Stokes parameters, $\xi_i$ are given in general by
\[
\xi_i=\f{\Phi_i}{\Phi_0},
\]
where $j=0,1,2,3$ and $\Phi_i$ is a function depending on the azimuthal 
angle, $\phi$ of the scattered photon beam when initial electron beam is taken 
along the $z$ axis:
\[
\Phi_j=\sum_{n=0}^4 \left(C_{jn}\,\cos n\phi+S_{jn}\,\sin n\phi\right),
\]
where $C_{jn}$ and $S_{jn}$ are certain real coefficients.
The circular polarization of the photon arising from Compton scattering, 
after averaging over the azimuthal
angle, is
\[
<\xi_2>=C_{20}/C_{00},
\]
where
\bea
C_{00}&=&\f{1}{1-y}+1-y-4\,r\,(1-r)-2\,\lae \ll\,r\,x\,(2r-1)\,(2-y)\non \\
C_{20}&=&2\,\lae\,r\,x\,\left[1+(1-y)\,(2r-1)^2\right]-
        \ll\,(2r-1)\,\left(\f{1}{1-y}+1-y\right).
\eea
A similar expression would hold for the circular polarization $<\o{\xi}_2>$ 
of the photon coming from the other beam.

\section{Density Matrix Elements for $\gamma\gamma\rightarrow 
t(\lambda_t)\overline{t}(\lambda_{\overline{t}})$}
\label{sec:rhoij}
 
Expressions for $\rho^{\pm}_{ij}$ used in 
eq.~(\ref{eq:csectiongg}) are given below.
\bea
\rho_{00}^+(\lt,\lt')&=&\f{1}{4}\,\sum_{\lg,\lg',\ltb}
\;M(\lg,\lg',\lt,\ltb)\,M^*(\lg,\lg',\lt',\ltb)\non\\
\rho_{22}^+(\lt,\lt')&=&\f{1}{4}\,\sum_{\lg,\ltb}
\;\left[M(\lg,\lg,\lt,\ltb)\,M^*(\lg,\lg,\lt',\ltb)\right.\non\\
&&-\left.\;M(\lg,-\lg,\lt,\ltb)\,M^*(\lg,-\lg,\lt',\ltb)\right]\non\\
\left(\rho_{20}^+(\lt,\lt')+\rho_{02}^+(\lt,\lt')\right)&=&
\f{1}{2}\,\sum_{\lg,\ltb}\lg\;M(\lg,\lg,\lt,\ltb)\,M^*(\lg,\lg,\lt',\ltb)\non\\
\left(\rho_{20}^+(\lt,\lt')-\rho_{02}^+(\lt,\lt')\right)&=&
\f{1}{2}\,\sum_{\lg,\ltb}\lg\;M(\lg,-\lg,\lt,\ltb)\,M^*(\lg,-\lg,\lt',\ltb)\non
\eea
\vskip .2cm
\bea
\rho_{00}^-(\ltb,\ltb')&=&\f{1}{4}\,\sum_{\lg,\lg',\lt}
\;M(\lg,\lg',\lt,\ltb)\,M^*(\lg,\lg',\lt,\ltb')\non\\
\rho_{22}^-(\ltb,\ltb')&=&\f{1}{4}\,\sum_{\lg,\lt}
\;\left[M(\lg,\lg,\lt,\ltb)\,M^*(\lg,\lg,\lt,\ltb')\right.\non\\
&&-\left.\;M(\lg,-\lg,\lt,\ltb)\,M^*(\lg,-\lg,\lt,\ltb')\right]\non\\
\left(\rho_{20}^-(\ltb,\ltb')+\rho_{02}^-(\ltb,\ltb')\right)&=&
\f{1}{2}\,\sum_{\lg,\lt}\lg\;M(\lg,\lg,\lt,\ltb)\,M^*(\lg,\lg,\lt,\ltb')\non\\
\left(\rho_{20}^-(\ltb,\ltb')-\rho_{02}^-(\ltb,\ltb')\right)&=&\f{1}{2}\,\sum_{\lg,\lt}\lg
\;M(\lg,-\lg,\lt,\ltb)\,M^*(\lg,-\lg,\lt,\ltb')\non
\eea
Substituting the amplitudes from eq.~(\ref{eq:ggttamp}) we get 
\bea
{\rho}_{00}^+(\pm,\pm)&=&{\rho}_{00}^-(\pm,\pm) \non \\
	&=&\f{1}{4}\,C\,
	\left\{ \f{s}{2\,m_t^2}\left[\left(1-\bt^4\right)
	+A\right]\pm \idt\,B_0\right\}
	\non \\
{\rho}_{00}^\pm(+-)&=&
-\f{1}{4}\,C\:\f{s^{3/2}\,\bt}{m_t^2}\,\idt\,\sin\tht\,
\cos\tht\,\left((1-\bt^2)+\bt^2\sin^2\tht\right)
\non\\[3mm]
{\rho}_{22}^+(\pm,\pm)&=&{\rho}_{22}^-(\pm,\pm) \non \\
	&=&\f{1}{4}\,C\,
	\left\{ \f{s}{2\,m_t^2}\left[\left(1-\bt^4\right)
	-A\right]\pm \idt\,B_2\right\}
	\non \\
{\rho}_{22}^\pm(+-)&=&
-\f{1}{4}\,C\:\f{s^{3/2}\,\bt}{m_t^2}\,
\idt\,\sin\tht\,\cos\tht\,
\left((1-\bt^2)-\bt^2\sin^2\tht\right)
\non\\[3mm]
{\rho}_{20}^\pm(\la,\la)&=&\f{1}{2}\,C\,\left\{\la \left(
	2\,\bt\pm D \right)
	+8m_t\,\idt\right\}\non\\
{\rho}_{20}^\pm(+-)&=&\mp\f{1}{2}\,C\,\sqrt{s}\,\left\{
	\f{\bt^2}{m_t}\,\sin^3\tht \right.\non \\
	&&\left.+\:\f{s\bt}{2\,m_t^2}\rdt\, \sin\tht
	\left((1-\bt^2)\,\cos\tht+\bt^2\,\sin^2\tht\right)
	\right\} \non\\[3mm]
{\rho}_{02}^\pm(\la,\la)&=&\f{1}{2}\,C\,\left\{\lambda\left(2\,\bt\mp
	D \right)
	+8m_t\,\idt\right\}\non\\
{\rho}_{02}^\pm(+-)&=&\pm\f{1}{2}\,C\,\sqrt{s}\,\left\{
	\f{\bt^2}{m_t}\,\sin^3\tht \right. \non \\
	&&\left.+\:\f{s\bt}{2\,m_t^2}\rdt\, \sin\tht
	\left((1-\bt^2)\,\cos\tht-\bt^2\,\sin^2\tht\right)
	\right\} \non
\eea
where $A,\,B_i,\,C$ and $D$ are given by
\bea
A&=&\bt^2\,\sin^2\tht\left[2-\bt^2\,\sin^2\tht\right]\non\\
B_0&=&\f{2\,s\bt}{m_t}\,\left[
	\left(2-\sin^2\tht\right)\left(
	1-\bt^2\right)-\bt^2\,\sin^4\tht\right] \non \\
B_2&=&\f{2\,s\bt}{m_t}\,\left[
	\left(2-\sin^2\tht\right)\left(
	1-\bt^2\right)+\bt^2\,\sin^4\tht\right] \non \\
C&=&16\,\pi^2\,Q_t^4\,\alpha^2\,
\f{16\, m_t^2}{s\,(1-\bt^2\cos^2\theta_t)^2} \non \\
D&=&\f{s\,\bt^2}{2\,m_t^2}\, 
	\sin^2\tht\cos\tht. \non
\eea
$Q_t = 2/3$ is the charge of the top quark. Note that we have kept only 
linear terms in $d_t$ assuming that the value of dipole form factor is
small and hence that higher order terms can be neglected.

\end{document}